\documentclass[aps,pra,preprint,superscriptaddress]{revtex4-1}

\usepackage{amsmath}
\usepackage{fixmath}
\usepackage{accents}
\usepackage{graphicx}%

\usepackage{hyperref}

\newcommand{\mb}{\mathbold}

\begin{document}
\title{Enhancement of the spontaneous emission in subwavelength quasi-two-dimensional waveguides and resonators}
\author{Mikhail Tokman}
\affiliation{Institute of Applied Physics, Russian Academy of Sciences}
\author{Zhongqu Long}
\affiliation{Department of Physics and Astronomy, Texas A\&M
University, College Station, TX, 77843 USA}
\author{Sultan AlMutairi}
\affiliation{Department of Physics and Astronomy, Texas A\&M
University, College Station, TX, 77843 USA}
\author{Yongrui Wang}
\affiliation{Department of Physics and Astronomy, Texas A\&M
University, College Station, TX, 77843 USA}
\author{Mikhail Belkin}
\affiliation{Department of Electrical and Computer Engineering, 
University of Texas at Austin, Austin, TX, 78712 USA}
\author{Alexey Belyanin}
\affiliation{Department of Physics and Astronomy, Texas A\&M
University, College Station, TX, 77843 USA}

\date{\today}

\begin{abstract} 

We consider a quantum-electrodynamic problem of the spontaneous emission from a two-dimensional (2D) emitter, such as a quantum well or a 2D semiconductor, placed in a quasi-2D waveguide or cavity with subwavelength confinement in one direction. We apply the Heisenberg-Langevin approach which includes dissipation and fluctuations in the electron ensemble and in the electromagnetic field of a cavity on equal footing. The Langevin noise operators that we introduce do not depend on any particular model of dissipative reservoir and can be applied to any dissipation mechanism. Moreover, our approach is applicable to nonequilibrium electron systems, e.g. in the presence of pumping, beyond the applicability of the standard fluctuation-dissipation theorem. We derive analytic results for simple but practically important geometries: strip lines and rectangular cavities. Our results show that a significant enhancement of the spontaneous emission, by a factor of order 100 or higher, is possible for quantum wells and other 2D emitters in a subwavelength cavity. 

\end{abstract}

\maketitle
\section{Introduction}
Enhancement of the radiative processes due to the localization of emitters in a subwavelength cavity (so-called Purcell enhancement \cite{purcell}) is a fundamental cavity-quantum electrodynamics (QED) effect which finds an increasingly broad range of applications in the areas as diverse as nanophotonics, plasmonics, linear and nonlinear optical sensing, and high-speed communications, to name a few. It has been studied theoretically and experimentally so many times that it is hard to believe that any further development is needed. However, there seems to be a significant gap in the formalism for the situations typically encountered in quantum optoelectronic devices, when the electron ensemble is out of equilibrium and there is strong dissipation both in the optical dipole oscillations in a macroscopic ensemble of fermionic emitters (e.g. electrons and holes in a semiconductor quantum well or a layer of quantum dots, or a 2D semiconductor such as MoS$_2$, or monolayer graphene) and for the electromagnetic (EM) field in a cavity. Examples include subwavelength semiconductor lasers \cite{fainman,chuang,ning,faist,nano4} and other  devices or circuits with subwavelength confinement in one or more dimensions e.g.~\cite{nano1,nano2,nano3}. In this case using a simple Purcell-type factor $\sim Q\lambda^3/V$, where $Q$ is a quality factor of EM modes in a cavity of volume $V$ and $\lambda$ is the emission wavelength, can drastically overestimate the cavity enhancement of the spontaneous emission. Although this fact is well known, a consistent QED theory including dissipation and fluctuations is usually replaced by a more phenomenological rate equations approach \cite{chuang}.  Recent theoretical analysis of subwavelength lasers \cite{fainman} did include QED Heisenberg-Langevin equations for the EM cavity modes, but not for the dynamics of the active medium.   

Here we use a consistent Heisenberg-Langevin approach \cite{et2015,et2017} which includes dissipation and fluctuations in the fermionic ensemble and in the EM field of a subwavelength cavity on equal footing. The Langevin noise operators that we introduce do not depend on any particular model of dissipative reservoir. Instead, they are derived directly from the condition of preserving the commutator for bosonic fields. Therefore, they can be applied to any dissipation/fluctuation mechanism. Moreover, our approach allows one to consider fluctuations due to nonequilibrium electron systems, e.g. in the presence of pumping, beyond the applicability of the standard fluctuation-dissipation theorem. 

We apply the general formalism to the problem of spontaneous emission  in a quasi-2D waveguide or cavity with subwavelength confinement in one direction. Remarkably, we are able to derive closed-form analytic results for all relevant quantities such as spontaneous emission power for simple but practically important geometries: strip lines and rectangular cavities. Our results provide general framework and convenient formulas for the evaluation of enhancement of linear and nonlinear radiative processes in such systems. Our results also indicate that a significant enhancement of the spontaneous emission, by a factor of order 100 or higher, is possible for  QWs and other 2D emitters sandwiched between metal plates in a subwavelength cavity. 

Section II describes the spatial structure of the EM field in a subwavelength quasi-2D electrodynamic structure and develops the quantization procedure. Section III introduces coupling to the fermionic system. Section IV derives and solves Heisenberg-Langevin equations for the density operator of quasiparticles and EM field operators. It also derives the expression for the spontaneous emission power and its useful limiting cases. 

\section{Electromagnetic field of a subwavelength cavity}
\subsection{Spatial structure of the EM field modes}

Consider a very thin layer of quantum dipole emitters (which we will call a quantum well (QW) for brevity, although it can be any fermionic system), placed inside a strip line or a cavity formed by two metallic planes at $ z= \pm L_z/2$ where $ L_z \ll c/ \sqrt{\bar{\varepsilon}}\omega$ , where $\bar{\varepsilon}$ is a typical (average) value of the dielectric constant
$\varepsilon=\varepsilon(z)$ of the filling; see Fig.~1.

\begin{figure}[h]
    \centering
    \includegraphics[width=0.8\columnwidth]{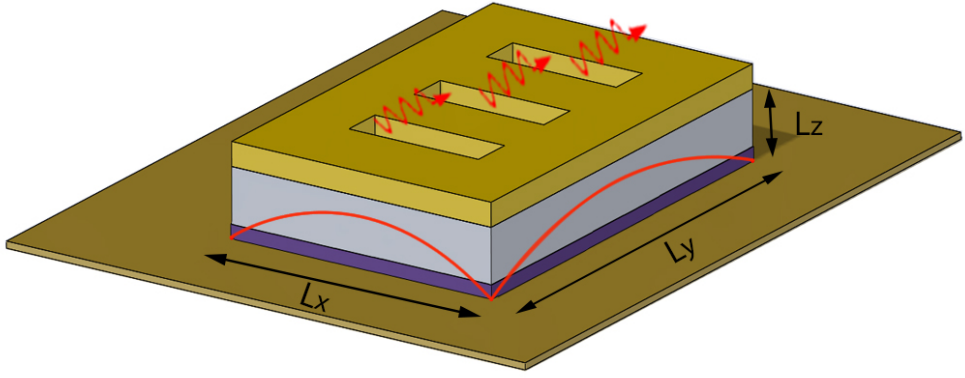}
    \caption{A sketch of a nanocavity with thickness $L_z$ much smaller than wavelength. An active layer of 2D emitters is shown in dark blue. The profile of the electric field  of the fundamental TE$_{011}$  mode is sketched on the sides. The radiation can be outcoupled through the gratings or cavity edges. }
	\label{fig1}
\end{figure}

A TM-polarized EM field is described by the following components of the electric field, magnetic
  field and electric induction:
\begin{equation}
\label{Eq:1}
(\mb{E}_{x,z},\mb{B}_y,\mb{D}_{x,z} )= {\rm Re} \left[ \left(\mb{\tilde{E}}_{x,z}(z),\mb{\tilde{B}}_y(z),\mb{\tilde{D}}_{x,z}(z)\right) e^{-i\omega t+iqx} \right]
\end{equation}
Where we assumed that the strip line is oriented along $x$. From Maxwell's equations,
\begin{equation}
\label{Eq:2}
\nabla \cdot \mb{D} =0,
\hspace{.5 cm} 
\nabla \times \mb{B} = \frac{\mb{\dot{D}}}{c},   
\hspace{.5 cm} 
\nabla \times \mb{E} = -\frac{\mb{\dot{B}}}{c}  
\end{equation}
together with the material equation,
\begin{equation}
\label{Eq:3}
\mb{D}=\varepsilon (z) \mb{E}
\end{equation}
we obtain:
\begin{equation}
\label{Eq:4}
\frac{\partial \tilde{D_z}}{\partial z} = -iq\tilde{D_x},
\hspace{.5 cm} 
 iq\tilde{B_y} =-\frac{i\omega \tilde{D_z}}{c},
 \hspace{.5 cm} 
 \frac{\partial \tilde{E_x}}{\partial z} = i \frac{\omega}{c	}\tilde{B_y}(z)+i\frac{q}{\varepsilon(z)} \tilde{D_z}
\end{equation}
The first equation in (4) yields
\begin{equation}
\tilde{D}_z=\tilde{D}_z \left(- \frac{L_z}{2} \right)- iq\int\limits_{-\frac{L_z}{2}}^z \tilde{D}_x dz' \nonumber
\end{equation}

For subwavelength thickness $L_z q \ll 1$ the previous equation gives $\tilde{D}_z \approx$ const, which corresponds to
the quasi-electrostatic structure of the field in the $(y, z)$ cross section of the strip line. From the second and third equations in Eq.~(\ref{Eq:4}) we can obtain

\begin{equation}
\label{Eq:5}
\frac{\partial \tilde{E}_x }{\partial z }=-i \frac{\omega^2 \tilde{D}_z }{qc^2}+i\frac{q}{\varepsilon(z)}\tilde{D}_z
\end{equation}
Next we integrate Eq.~(\ref{Eq:5}) as $\displaystyle \int_{- \frac{L_z}{2}}^{ \frac{L_z}{2}} dz\dotsc$ , taking into account $\tilde{D}_z \approx const$ and the boundary conditions
on the metal planes: $\tilde{E}_x(+\frac{L_z}{2})=\tilde{E}_x(- \frac{L_z}{2})
=0$. As a result, we obtain the dispersion relation:
\begin{equation}
\label{Eq:6}
\frac{\omega^2}{q^2c^2}=\frac{1}{L_z}\int\limits_{- \frac{L_z}{2}}^{ \frac{L_z}{2}}\frac{dz}{\varepsilon(z)}.
\end{equation}
Since the direction of $x$-axis was arbitrary, we can represent the electric field vector as
\begin{equation}
\label{Eq:7}
\mb{E}=D_q\mb{F_q}(\mb{r})e^{-i\omega _qt} + \mathrm{C.C.},
\end{equation}
where the factor $\mb{F_q}(\mb{r})$ determines the spatial structure of the field:
\begin{equation}
\label{Eq:8}
\mb{F_q}(\mb{r})=\mb{z_0}\frac{e^{i\mb{qr}}}{\varepsilon(z)},
\end{equation}
vector $\mb{q}$ is in the $(x, y)$ plane, $D_q$ is a constant which in this case corresponds to a $z$-independent amplitude of the electric induction.
According to the Brillouin concept, one can use the waves defined by Eqs.~(\ref{Eq:6})-(\ref{Eq:8}) to construct any waveguide and cavity modes. They have quasi-TEM polarization. In particular, if the sides $y =\pm L_y/ 2$ are also metal-coated, consider the lowest order (01) waveguide mode:
\begin{equation}
\label{Eq:9}
\mb{E}=D_{q_x}\mb{F}_{q_x}(\mb{r})e^{-i\omega _{q_x}t} + \mathrm{C.C.},
\hspace{.5 cm} 
q^2_x+\left(\frac{\pi}{L_y}\right)^2 = \frac{\omega^2}{c^2}\frac{L_z}{\int\limits_{- \frac{L_z}{2}}^{ \frac{L_z}{2}}\varepsilon(z)^{-1}dz}
\end{equation}
where the explicit form to the factor $\mb{F_{q_x}}(\mb{r}) \propto e^{-i q_x x}$ is given below. If the facets $x = \pm L_x/ 2$ are metal-coated as well, the waveguide becomes a resonator and the lowest order modes are $TE_{01N}$:
\begin{equation}
\label{Eq:10}
\mb{E}=D_N\mb{F}_N(\mb{r})e^{-i\omega_N t} + \mathrm{C.C.},
\hspace{.5 cm} 
\left(\frac{N\pi}{L_x}\right)^2+\left(\frac{\pi}{L_y}\right)^2 = \frac{\omega^2}{c^2}\frac{L_z}{ \int\limits_{- \frac{L_z}{2}}^{+\frac{L_z}{2}}\varepsilon(z)^{-1}dz}
\end{equation}
In Eqs.~(\ref{Eq:9}) and (\ref{Eq:10}) the factors $D_{q_x}$ and $D_N$ are coordinate-independent amplitudes of the electric induction. The factors $\mb{F}_{\mb{q},q_x,N} (\mb{r}$) in Eqs.~(\ref{Eq:7}), (\ref{Eq:9}), (\ref{Eq:10}) can be written in the same form using the index $\nu=\mb{q}, q_x, N $ to denote a corresponding spatial structure:
\begin{equation}
\label{Eq:11}
\mb{F}_\nu(\mb{r})=\mb{z_0}\frac{\zeta_\nu(x,y)}{\varepsilon(z)},  
\hspace{.2 cm} 
\zeta_\mb{q}=e^{i\mb{qr}},  
\hspace{.2 cm} 
\zeta_{q_x}=\cos \left( \frac{\pi y}{L_y} \right) e^{iq_x x},  
\hspace{.2 cm} 
\zeta_N=\cos \left(\frac{\pi y}{L_y}\right) \times \left\{  
\begin{matrix}
\cos\left(\displaystyle\frac{N_{odd}\pi x}{L_x}\right) \\ \sin \left(\displaystyle\frac{N_{even}\pi x}{L_x}\right)
\end{matrix} 
\right.
\end{equation}
where $\int_S \zeta_\nu \zeta^*_{\nu'} d^2r \propto \delta_{\nu\nu'}$ . For a particular case of a uniform dielectric constant, Eqs.~(\ref{Eq:6})-(\ref{Eq:11}) are exact.
Similar equations can be derived if one simply utilizes jumps of the dielectric constants on the sides instead of metal coating. Even without any jump in the dielectric constants, an open end of a thin waveguide with vertical size much smaller than wavelength is  a good reflector and therefore any radiation losses through the facets are small and are not affecting the mode spatial structure significantly.

\subsection{Field quantization in a subwavelength waveguide/cavity}

Here we consider field quantization in a volume $V = L_z S$ , where $S = L_x \times L_y$ . The field operator can be represented in a standard form \cite{fain,SZ}:
\begin{equation}
\label{Eq:12}
\hat{\mb{E}}= \sum_\nu [\mb{E(r)}_\nu\hat{c}_\nu + \mb{E^*(r)}_\nu\hat{c_\nu}^\dagger]
\end{equation}
where $\hat{c}_\nu$ and $\hat{c_\nu}^\dagger$ are boson annihilation and creation operators, $\displaystyle \mb{E}_\nu(\mb{r})=\mb{z_0}\frac{\zeta_\nu(x,y)}{\varepsilon(z)} D_\nu $, and $D_\nu$ is the normalization constant corresponding to the $z$-independent amplitude of the electric induction. The value of $D_\nu$ needs to be chosen in such a way that the commutation relation for boson operators $\hat{c}_\nu$ and $ \hat{c_\nu}^\dagger$ have a standard form $[\hat{c}_\nu,\hat{c_\nu}^\dagger]=\delta_{\nu\nu'}$. In this case the field Hamiltonian will also be standard:

\begin{equation}
\label{Eq:13}
\hat{H}_f=\sum_\nu \hbar\omega_\nu(\hat{c_\nu}^\dagger\hat{c}_\nu +\frac{1}{2})
\end{equation}

To find the explicit expression for $ D_\nu$ we apply the phenomenological procedure of field quantization in a medium \cite{fain,ginzburg} which was justified in \cite{vdovin} based on a rigorous quantum electrodynamics theory. According to this approach, the normalization is determined by the requirement that the classical energy density $W$ of the EM field $\mb{E}=\mb{E_\nu(r)}e^{-i\omega_\nu t} + \mathrm{C.C}.$, $\mb{B}=\mb{B_\nu(r)}e^{-i\omega_\nu t} +\mathrm{C.C} $ give the total energy of $ \int_V Wd^3r= \hbar \omega_\nu$. For our strip line this procedure yields the following expression for the normalization constant (see Appendix A):
\begin{equation}
\label{Eq:14}
|D_\nu |^2= \frac{2\pi\hbar\omega_\nu}{\int_S \zeta_\nu \zeta^*_{\nu} d^2r  \times \int\limits_{- \frac{L_z}{2}}^{ \frac{L_z}{2}}\displaystyle \frac{1}{2\varepsilon^2(\omega_\nu,z)\omega_\nu}\left[\frac{\partial(\omega^2 \varepsilon(\omega,z)}{\partial\omega}\right] _{\omega=\omega_\nu} dz },
\end{equation}
where $\int_S \zeta_\mb{q} \zeta^*_\mb{q} d^2r =S$, $\int_S \zeta_{q_x} \zeta^*_{q_x} d^2r =S/2 $ and $\int_S \zeta_N \zeta^*_N d^2r =S/4$. In the limiting case of plane waves in a homogeneous medium Eq.~(\ref{Eq:14}) corresponds to a standard normalization of the electric field
\cite{fain,vdovin,ginzburg}; indeed, taking into account that in a homogeneous medium $D_\nu =E_\nu \varepsilon(\omega_\nu),$ Eq.~(\ref{Eq:14}) gives $|E_\nu |^2=\displaystyle\frac{2\pi \hbar\omega_\nu}{\displaystyle\frac{V}{2\omega_\nu}\left[\frac{\partial(\omega^2 \varepsilon(\omega,z)}{\partial\omega} \right] _{\omega=\omega_\nu}} $, where $V = L_z S $ is the quantization volume.

\section{Non-dissipative dynamics of a coupled system of photons and electrons}
\subsection{ General formalism}
We will denote a quantum state of an electron in a QW or any other 2D nanostructure by a band index $m$ which may include also the subband, spin, and valley index as needed, and the 2D quasimomentum $\mb{k}$ corresponding to the motion in $(x, y)$ plane. The second-quantized energy of a system of such quasiparticles is
\begin{equation}
\label{Eq:15}
\hat{H}_e = \sum_{m\mb{k}} W_{m\mb{k}}\hat{a}^\dagger_{m\mb{k}} \hat{a}_{m\mb{k}}
\end{equation}
where $\hat{a}^\dagger_{m\mb{k}} , \hat{a}_{m\mb{k}} $ are creation and annihilation operators of fermions, $W_{m\mb{k}}  \equiv W_{mm\mb{k}\mb{k}}$ are the diagonal matrix elements of the energy operator of a quasiparticle. The eigenfunctions can be written as
\begin{equation}
\label{Eq:16}
| m,\mb{k}\rangle = \frac{e^{i \mb{kr}}}{\sqrt{S}}\psi_m(z)
\end{equation}
where $ \int_S e^{i(\mb{k}-\mb{ k'})\mb{r}}d^2 r=S \delta_{\mb{k}\mb{k}'}$,  $\int\limits_{- \frac{l}{2}}^{ \frac{l}{2}} \psi_m(z) \psi_n^*(z) dz=\delta_{mn}$.
Here we assume that a 2D nanostructure occupies a region $-\l/2 \leq z \leq l/2$, $l \leq L_z$. The total Hamiltonian of a coupled system of photons and electrons is
\begin{equation}
\label{Eq:17}
\hat{H}=\hat{H}_f+\hat{H}_e+\hat{V}
\end{equation}
where the operators $\hat{H}_f$ and $\hat{H}_e$ are given by Eqs.~(\ref{Eq:13}) and (\ref{Eq:15}), and $\hat{V}$ is the interaction Hamiltonian, which can also be written in the second-quantized form:
\begin{equation}
\label{Eq:18}
\hat{V}=\sum_{mn\mb{k}\mb{k}'} \hat{V}_{nm\mb{k}'\mb{k}}\hat{\rho}_{nm\mb{k}\mb{k}'}
\end{equation}
where $\hat{\rho}_{nm\mb{k}\mb{k}'}=\hat{a}^\dagger_{n\mb{k}'}\hat{a}_{m\mb{k}}$ is the density operator. Matrix elements $\hat{V}_{nm\mb{k}'\mb{k}}$ in Eq.~(\ref{Eq:18}) are operators since they
depend on the quantum field.

Taking into account the quasi-electrostatic structure of the electric field in the transverse cross-section of a strip line, we can write the interaction Hamiltonian in the electric potential approximation:
\begin{equation}
\label{Eq:19}
\hat{V}=e \int\limits_{-l/2}^z \hat{E}_z dz
\end{equation}
Using Eq.~(\ref{Eq:12}) for the field operator, the matrix elements of the interaction Hamiltonian are
\begin{equation}
\label{Eq:20}
\hat{V}_{nm\mb{k}'\mb{k}} = -\tilde{d}_{nm} \sum_\nu (D_\nu \hat{c}_\nu\zeta_{\mb{k}'\mb{k}}^{(\nu)} + D^*_\nu \hat{c}^\dagger_\nu\zeta_{\mb{k}'\mb{k}}^{(\nu)\dagger})
\end{equation}
where $\tilde{d}_{nm}$ is the effective dipole moment of the optical transition:
\begin{align}
\label{Eq:21}
\tilde{d}_{nm} &= -e \int\limits_{-l/2}^{l/2} \left[\psi_n^*(z)\left(\int\limits_{-l/2}^z \frac{dz'}{\varepsilon(z')}\right)\psi_m(z)\right]dz \\
\label{Eq:22}
\zeta_{\mb{k}'\mb{k}}^{(\nu)} &= \frac{1}{S}\int_S e^{-i\mb{k}'r} \zeta_\nu(x,y)e^{i\mb{k}r} d^2r, 
\zeta_{\mb{k}'\mb{k}}^{(\nu)\dagger}=(\zeta_{\mb{k}'\mb{k}}^{(\nu)})^* 
\end{align}
For a homogeneous medium, in which $E_\nu = D_\nu/\varepsilon$ , Eq.~(20) will contain a standard expression
 $\tilde{d}_{nm}D_\nu = -e \langle n | z | m\rangle E_\nu$.
 
 The Hamiltonian Eq.~(\ref{Eq:17}) gives rise to the Heisenberg equations for photon operators:
\begin{align}
\label{Eq:23}
\dot{\hat{c}}_\nu &= \frac{i}{\hbar}[\hat{H},\hat{c}_\nu ]= -i\omega_\nu \hat{c}_\nu + \frac{i}{\hbar}D^*_\nu \sum_{mn\mb{k}\mb{k}'} \tilde{d}_{nm} \zeta_{\mb{k}'\mb{k}}^{(\nu)\dagger} \hat{\rho}_{mn\mb{k}\mb{k}'},\nonumber \\
 \dot{\hat{c}}^\dagger_\nu &= \frac{i}{\hbar}[\hat{H},\hat{c}^\dagger_\nu ]= i\omega_\nu \hat{c}^\dagger_\nu - \frac{i}{\hbar}D_\nu \sum_{mn\mb{k}\mb{k}'} \tilde{d}_{nm} \zeta_{\mb{k}'\mb{k}}^{(\nu)} \hat{\rho}_{mn\mb{k}\mb{k}'}
\end{align}
We write a similar equation for the density operator using a shortcut notation $ |m,\mb{k}\rangle = |\mu \rangle $ for brevity. Using the fundamental commutation relation \cite{vdovin, te2013, belyanin2013}
\begin{equation}
\label{Eq:24}
[\hat{\rho}_{\mu'\eta'},\hat{\rho}_{\mu\eta}] = (\delta_{\mu'\eta} \hat{\rho}_{\mu\eta'}-\delta_{\mu\eta'} \hat{\rho}_{\mu'\eta})
\end{equation}
which is valid whether the creation and annihilation operators $\hat{a}_\eta^\dagger$ and $\hat{a}_\mu$ satisfy the commutation relations for bosons or fermions, we obtain:
\begin{equation}
\label{Eq:25}
\dot{\hat{\rho}}_{\mu\eta} = \frac{i}{\hbar}[\hat{H},\hat{\rho}_{\mu\eta} ]= - \frac{i}{\hbar} \sum _{\mu'} (\hat{H}_{\mu\mu'} \hat{\rho}_{\mu'\eta} -\hat{\rho}_{\mu\mu'}\hat{H}_{\mu'\eta} )
\end{equation}
The resulting equation for the density operator has the same form as the von Neumann equation, although the original Heisenberg equation had an opposite sign in front of the commutator \cite{vdovin, te2013,belyanin2013}. This is to be expected, because for time-dependent Heisenberg operators $\hat{a}_\eta^\dagger$ and $\hat{a}_\mu$
the average of dyadics $ \hat{\rho}_{\mu\eta}= \hat{a}_\eta^\dagger \hat{a}_\mu $ over the initial quantum state should correspond to a usual density matrix.

\subsection{Matrix elements of the interaction Hamiltonian}

The form of the interaction Hamiltonian for the fields with different spatial structure depends on the matrix elements $ \zeta_{\mb{k}'\mb{k}}^{(\nu)}$ defined in Eq.~(\ref{Eq:22}). In particular, for plane waves we obtain $ \zeta_{\mb{k}'\mb{k}}^{(\mb{q})}=\delta_{\mb{k}';\mb{k}+\mb{q}} $. For a waveguide or a cavity the corresponding expressions for $\zeta_{\mb{k}'\mb{k}}^{(q_x)}$ and $\zeta_{\mb{k}'\mb{k}}^{(N)}$ are quite cumbersome and are given in Appendix B.

If we take into account that the de Broglie wavelength of electrons is typically much smaller than the spatial scale of the EM field, i.e. $  k\gg|\mb{q}| , q_x , \displaystyle\frac{\pi N}{L_x} , \displaystyle\frac{\pi N}{L_y} ,$ the expressions for matrix elements are simplified. Indeed, in this case we can assume that the optical transitions are direct in momentum space and take $  \zeta_{\mb{k}'\mb{k}}^{(\nu)}\approx \alpha_\nu \delta_{\mb{k}'\mb{k}}$ . The factor in front of the delta-function is one for plane waves; for a waveguide or a cavity one should choose $\alpha_\nu = \sqrt{\sum_{\mb{k}'}  \zeta_{\mb{k}'\mb{k}}^{(\nu)} \zeta_{\mb{k}\mb{k}'}^{(\nu)  \dagger}} $. With this choice, a resonance line which is “smeared” in the quasimomentum space can be reduced to the delta-function $\alpha_\nu\delta_{\mb{k}'\mb{k}} $ while conserving the sum of intensities of all transitions within the line. The Parseval theorem then gives $\sum_{\mb{k}'}  \zeta_{\mb{k}'\mb{k}}^{(\nu)} \zeta_{\mb{k}\mb{k}'}^{(\nu)  \dagger} = S^{-1} \int_S \zeta_\nu \zeta_\nu^* d^2r$ (see Appendix B). As a result the matrix element can be written in the same form for plane waves, in a waveguide,
and in a cavity:
\begin{equation}
\label{Eq:26}
\hat{V}_{nm\mb{k}\mb{k}'} \approx -\tilde{d}_{nm} \sum_\nu (\tilde{D}_\nu \hat{c}_\nu +\tilde{D}^*_\nu \hat{c}^\dagger_\nu)\delta_{\mb{k}'\mb{k}}
\end{equation}
where
\begin{align}
\label{Eq:27}
| \tilde{D}_\nu|^2 &= \frac{2\pi \hbar\omega_\nu}{SG(L_z,\omega_\nu)}  \\
\label{Eq:28}
G(L_z,\omega_\nu)  &=  \int\limits_{- \frac{L_z}{2}}^{ \frac{L_z}{2}}\frac{1}{2\varepsilon^2(\omega_\nu,z)\omega_\nu} \left[ \frac{\partial(\omega^2 \varepsilon(\omega,z)}{\partial\omega} \right]_{\omega=\omega_\nu}dz 
\end{align}
Note that in a uniform nondispersive medium $\tilde{d}_{mn} = d_{mn}/\varepsilon$ and $G = L_z/\varepsilon$.

\subsection{The probability of the spontaneous emission}

Consider a spontaneous radiative transition $m\to n$ for a quasiparticle in an open electrodynamic system, e.g. in the space between two conducting planes or in a waveguide. The transition probability is usually calculated using Fermi's golden rule \cite{LL-QM}:
\begin{equation}
\label{Eq:29}
A_{m\to n} = \frac{2\pi}{\hbar^2}\int d\Pi_f |V_{fi}|^2 \delta\left(\frac{W_i}{\hbar}-\frac{W_f}{\hbar}-\omega_\nu\right)
\end{equation}
where the integration $\int d\Pi_f$ is taken over all final states of a system labeled by $f$. The matrix element $  V_{fi}$ in this case is equal to $\langle 1_\nu | \hat{V}_{nm\mb{k}'\mb{k}}|0_\nu\rangle$, where $|n_\nu\rangle$ is a Fock state of photons. Using Eqs.~(\ref{Eq:20}) and (\ref{Eq:26})-(\ref{Eq:28}) we obtain
\begin{equation}
\label{Eq:30}
V_{fi}=  -\tilde{d}_{nm}  D^*_\nu \zeta_{\mb{k}'\mb{k}}^{(\nu )\dagger} \approx  -\tilde{d}_{nm}  \tilde{D}^*_\nu \delta_{\mb{k}'\mb{k}}
\end{equation}
Taking into account the photon density of states, one can get for the radiation emitted into space between two conducting planes
\begin{equation}
d\Pi_f = \frac{S|\mb{q}| d\theta d\omega_q}{(2\pi)^2 |\partial\omega_q/\partial \mb{q}|} \nonumber
\end{equation}
where $\theta$ determines the direction of vector $\mb{q}$ in the $(x,y)$ plane. For the radiation emitted into a waveguide,
\begin{equation}
d\Pi_f = \frac{L_x\omega_{q_x}}{2\pi |\partial\omega_{q_x}/\partial q_x|} \nonumber
\end{equation}
The resulting expressions for the spontaneous emission probabilities are
\begin{equation}
\label{Eq:31}
A_{m\to n}^{(\mb{q})} = \frac{2\pi |\tilde{d}_{mn}|^2 \omega_{mn}|\mb{q}|}{\hbar |\partial\omega_q/\partial \mb{q}|_{\displaystyle\omega_q =\omega_{mn}}G(L_z,\omega_\nu)}
\end{equation}
\begin{equation}
\label{Eq:32}
A_{m\to n}^{(q_x)} = \frac{2\pi |\tilde{d}_{mn}|^2 \omega_{mn}}{\hbar |\partial\omega_{q_x}/\partial q_x|_{\displaystyle\omega_{(q_x)} =\omega_{mn}} L_y G(L_z,\omega_\nu)}
\end{equation}
where $\omega_{mn}$ is the transition frequency.

In order to use Fermi's golden rule in a cavity, one has to formally introduce the density of states assuming that the modal spectrum is spread near the resonance frequency $\omega_{mn}$  by the linewidth
$\Delta \omega$  :
\begin{equation}
\label{Eq:33}
d\Pi_f = \frac{(\Delta\omega/2\pi)}{(\omega_{mn}-\omega_N)^2+(\Delta\omega/2)^2}d\omega
\end{equation}
which results in
\begin{equation}
\label{Eq:34}
A_{m\to n}^{(N)}= \frac{2\pi |\tilde{d}_{mn}|^2 \left(\displaystyle \frac{4\omega_{mn}}{\Delta \omega}\right)} {\hbar L_x L_yG(L_z,\omega_\nu)}
\end{equation}
Eq.~(\ref{Eq:34}) is also valid for a waveguide at a critical frequency, i.e. for $  |\partial\omega_{q_x}/\partial q_x|_{\displaystyle\omega_{(q_x)} =\omega_{mn}}=0$, because such a system is effectively a cavity. In a homogeneous medium, expressions (\ref{Eq:31}), (\ref{Eq:32}) and (\ref{Eq:34}) can be simplified. In this case Eqs.~(\ref{Eq:21}) and (\ref{Eq:28}) lead to
\begin{equation}
\frac{|\tilde{d}_{mn}|^2}{G(L_z,\omega_\nu)}=\frac{|d_{mn}|^2}{\displaystyle\frac{L_z}{2\omega_\nu}\left[\frac{\partial(\omega^2\varepsilon)}{\partial \omega}\right]_{\omega=\omega_\nu}} \nonumber
\end{equation}
Finally we compare the spontaneous emission probability in a cavity with that in free space. The latter is equal to $\displaystyle A^{(0)} = \frac{4\omega^3|d_{mn}|^2 \sqrt{\varepsilon} }{3\hbar c^3}$. Their ratio is
\begin{equation}
\label{Eq:35}
\frac{A_{m\to n}^{(N)}}{A^{(0)}} \approx\frac{3\pi}{2}\frac{(c/\omega\sqrt{\varepsilon})^3}{L_x L_y L_z}\left(\frac{4\omega_{21}}{\Delta \omega}\right)
\end{equation}
Note that in Eq.~(\ref{Eq:35}) the minimal lateral sizes of an electrodynamic system we consider are $L_{x,y}=\pi c/\omega\sqrt{\varepsilon}$ , whereas the value of $ L_z$ can be much smaller.

Up to a numerical factor which depends on geometry, Eq.~(\ref{Eq:35}) is a widely used expression for the Purcell enhancement of the spontaneous emission. However, Eqs.~(\ref{Eq:31}), (\ref{Eq:32}), and (\ref{Eq:34}) do not include the effects of nonradiative relaxation in an ensemble of fermions. Moreover, the above approach does not allow one to determine the line broadening in a cavity in a consistent way. To include all dissipation processes consistently, we use the Heisenberg-Langevin formalism.

\section{Dissipative dynamics in an ensemble of photons and electrons}

\subsection{Heisenberg-Langevin equations for the quasiparticle density operator}

Dissipative effects in an open quantum system can be taken into account by adding the relaxation operator $\hat{R}_{\mu\eta}$ and corresponding Langevin noise operator $\hat{F}_{\mu\eta}$ to the right-hand side of Eq.~(\ref{Eq:25})\cite{SZ,et2015,et2017,david,te2013,belyanin2013}. One cannot add dissipation phenomenologically, without including Langevin sources, because this would violate the fundamental commutation relation Eq.~(\ref{Eq:24}) \cite{et2015,et2017,te2013,belyanin2013}. For the simplest model of ``transverse'' relaxation, when
\begin{equation}
\label{Eq:36}
\hat{R}_{\mu\neq\eta} = -\gamma_{\mu\eta} \hat{\rho}_{\mu\eta}.
\end{equation}
Refs.~\cite{et2015,et2017} derived the following expressions for the commutator and correlator of the Langevin noise (for a particular case of a two-level system):

\begin{equation}
\label{Eq:37}
\begin{matrix} 
[\hat{F}_{\mu\eta}(t'),\hat{F}^\dagger_{\mu\eta}(t)] = ( -\gamma_{\mu\eta}( \hat{\rho}_{\eta\eta}- \hat{\rho}_{\mu\mu}) +\hat{R}_{\eta\eta} - \hat{R}_{\mu\mu})\delta(t'-t) \\
\langle \hat{F}^\dagger_{\mu\eta}(t),\hat{F}_{\mu\eta}(t') \rangle = (2\gamma_{\mu\eta} \langle\hat{\rho}_{\mu\mu}\rangle+\langle \hat{R}_{\mu\mu}\rangle)\delta (t'-t)  
\end{matrix}
\end{equation}

where $\hat{F}^\dagger_{\mu\eta}=\hat{F}_{\eta\mu}$ and the symbol $\langle\dotsc \rangle$ means in this case the averaging over both the initial quantum state
and the statistics of a dissipative reservoir. The dissipation operator in its simplest form of Eq.~(\ref{Eq:36}) implies the absence of any inertia in a dissipative subsystem; that is why the noise operator turns out to be delta-correlated in time. Note that for degenerate fermion distributions Eqs.~(\ref{Eq:37}) are valid if the evolution equation for the density operator includes exchange effects which take care of Pauli blocking.

The nonzero value of the relaxation operator for populations, $\hat{R}_{\mu\mu} \neq 0$ in Eq.~(\ref{Eq:37}) corresponds to the nonequilibrium distribution. A steady-state distribution can be nonequilibrium because of an external pumping. An incoherent pumping generally redistributes populations over many subbands; therefore within the model taking into account a limited number of subbands such a pumping is convenient to introduce as a source $ \hat{J}_{\mu\eta}$ in the evolution equation for the density operator. This way we can assume that there is a “generalized” relaxation operator $\hat{\tilde{R}}_{\mu\eta}= \hat{R}_{\mu\eta}+\hat{J}_{\mu\eta}$ on the right-hand side of Eq.~(\ref{Eq:25}), and the steady-state (but not necessarily equilibrium) distribution corresponds to the condition $\langle \hat{\tilde{R}}_{\mu\mu}\rangle=0$ for all $\mu$. Of course, the modification of the relaxation operator causes the noise operator to change. However, within the simplest model of Eq.~(\ref{Eq:36}) this does not affect the general form of Eqs.~(\ref{Eq:37}). One just needs to keep in mind that the relaxation constants $\gamma_{\mu\eta}$ and operators $\hat{R}_{\mu\mu} $ in Eq.~(\ref{Eq:36}), (\ref{Eq:37}) contain the contribution from incoherent pumping.

The equation for the density operator can be further simplified if we (i) include only two subbands, i.e. $m,n = 1,2$; (ii) assume that optical transitions in the interaction Hamiltonian are direct; see Eq.~(\ref{Eq:26}). In this case the equation for the off-diagonal density operator elements includes only the elements  $\hat{\rho}_{21\mb{kk}}$ and $\hat{\rho}_{12\mb{kk}} = \hat{\rho}^\dagger_{21\mb{kk}}$ Finally, (iii) we assume populations to satisfy $\hat{R}_{11\mb{k}\mb{k}}=\hat{R}_{22\mb{k}\mb{k}}=0$. This gives
\begin{equation}
\label{Eq:38}
\dot{\hat{\rho}}_{21\mb{\mb{k}\mb{k}}}+i\omega_{21}(\mb{k})\hat{\rho}_{21\mb{kk}} +\gamma_{21\mb{kk}} \hat{\rho}_{21\mb{kk}} = \frac{i\tilde{d}_{21}}{\hbar} \left( \sum_\nu \tilde{D}_\nu \hat{c}_\nu\right)\cdot (\hat{\rho}_{11\mb{kk}}-\hat{\rho}_{22\mb{kk}}) +\hat{F}_{21\mb{kk}},
\end{equation}
where $\omega_{21}(\mb{k})=\displaystyle\frac{W_{2\mb{k}}-W_{1\mb{k}}}{\hbar}$.

As usual, the properties pf the Langevin source $\hat{F}_{21\mb{kk}}(t) $ in Eq.~(\ref{Eq:38}) are convenient to express  through the properties of its spectral components: $\hat{F}_{21\mb{kk}}(t) = \underaccent{\infty}{\int} \hat{F}_{\omega;21\mb{kk}} e^{i\omega t} d\omega, \hat{F}_{-\omega;12\mb{kk}}=\hat{F}^\dagger_{\omega;21\mb{kk}}$ . Taking into account that $\hat{R}_{11\mb{k}\mb{k}}=\hat{R}_{22\mb{k}\mb{k}}=0$, we can get from Eq.~(\ref{Eq:37}) (see also \cite{et2015,belyanin2013})
\begin{equation}
\label{Eq:39}
\langle \hat{F}^\dagger_{\omega;21\mb{kk}} \hat{F}_{\omega ';21\mb{kk}}\rangle = \frac{\gamma_{21\mb{kk}}}{\pi}n_{2\mb{k}} \delta(\omega-\omega'), 
\hspace{.5 cm}
\langle \hat{F}_{\omega;21\mb{kk}} \hat{F}^\dagger_{\omega';21\mb{kk}}\rangle = \frac{\gamma_{21\mb{kk}}}{\pi}n_{1\mb{k}} \delta(\omega-\omega'),
\end{equation}
where $n_{1\mb{k}}=\langle\hat{\rho}_{11\mb{kk}}\rangle $ and $n_{2\mb{k}}=\langle\hat{\rho}_{22\mb{kk}}\rangle$  are constant populations supported by pumping.

\subsection{Heisenberg-Langevin equations for field operators}

Similarly to relaxation in the medium, relaxation of the EM field gives rise to the noise sources in the equations for field operators \cite{SZ}. When field absorption by fermions is included, the noise term for the EM field appears due to Langevin noise terms in the density operator equations \cite{et2015,et2017,belyanin2013,tokman2015}. Including any additional field absorption unrelated to absorption in the medium should be accompanied by adding Langevin noise terms directly to field equations. We take into account this additional absorption for the $\nu$th mode of the field by including phenomenological dissipative operators $-\Gamma \hat{c}_\nu$ and $-\Gamma \hat{c}_\nu^\dagger$ to the right-hand side of the field equations (\ref{Eq:23}). To preserve the commutation relation $[\hat{c}_\nu,\hat{c}_\nu^\dagger]$ we need to add the Langevin noise operator $\hat{L}(t)$, satisfying the commutation relation $ [\hat{L}(t'),\hat{L}^\dagger(t)]= 2\Gamma \delta(t-t') $ (see Appendix C). Its correlator is equal to $\langle\hat{L}^\dagger(t') \hat{L}(t)\rangle= \Xi \cdot2\Gamma \delta(t-t')$, where the parameter $\Xi $ is determined by a state of a dissipative resevoir. When the latter is in equilibrium, we obtain \cite{SZ} $\Xi = (e^{\hbar \omega_\nu/T}-1)^{-1}$.

Next we take into account that the dissipation of a given $\nu$th mode of the EM field could also be due to absorption in metal walls and bulk material unrelated to the active medium. In this case we add the dissipative operators to the right-hand side of Eq.~(\ref{Eq:23}), $-(\Gamma_r + \Gamma_\sigma)\hat{c}_\nu $ and $ -(\Gamma_r + \Gamma_\sigma)\hat{c}^\dagger_\nu $, together with corresponding Langevin noise terms, $\hat{L}_r^{(\nu)} $ and $\hat{L}^{(\nu)}_\sigma$ . Here the factor $\Gamma_r $ describes radiative and diffraction losses out from the cavity and $\Gamma_\sigma $ describes Ohmic losses. Taking into account Eq.~(\ref{Eq:26}) for the interaction Hamiltonian, we obtain
\begin{equation}
\label{Eq:40}
\dot{\hat{c}}_\nu + (i\omega_\nu +\Gamma_r+\Gamma_\sigma)\cdot\hat{c}_\nu=\frac{i\tilde{d}_{12}\tilde{D}^*_\nu}{\hbar}\sum_\nu \hat{\rho}_{21\mb{kk}} +\hat{L}_r^{(\nu)}+\hat{L}^{(\nu)}_\sigma .
\end{equation}
Here the Langevin sources can again be defined through the properties of their spectral components:
\begin{gather}
\hat{L}^{(\nu)}_{r,\sigma} =\underaccent{\infty}{\int} \hat{L}^{(\nu)}_{r,\sigma;\omega}e^{-i\omega t}d\omega, 
\hspace{.5 cm} \hat{L}^{(\nu)}_{r,\sigma;-\omega}=\hat{L}^{(\nu)\dagger}_{r,\sigma;\omega} ; 
 \nonumber
\\
\langle \hat{L}^{(\nu')\dagger}_{r,\sigma';\omega'}  \hat{L}^{(\nu)}_{r,\sigma;\omega} \rangle = n_{T_{r,\sigma}}(\omega_\nu)\frac{\Gamma_{r,\sigma}\delta_{\nu\nu'}}{\pi}	\delta(\omega-\omega'),\nonumber \\
\label{Eq:41}
 \langle \hat{L}^{(\nu)}_{r,\sigma;\omega}  \hat{L}^{(\nu'\dagger)}_{r,\sigma';\omega'} \rangle = [n_{T_{r,\sigma}}(\omega_\nu)+1]\frac{\Gamma_{r,\sigma}\delta_{\nu\nu'}}{\pi}	\delta(\omega-\omega'),
\end{gather}
where $n_{T_{r,\sigma}}(\omega_\nu) = \displaystyle\frac{1}{e^{\hbar\omega_\nu/T_{r,\sigma}}-1}$ and $T_{r,\sigma}$ are the temperature of the ambient space which controls radiative losses and the bulk material inside the cavity. The presence $\delta_{\nu\nu'}$ in Eq.~(\ref{Eq:41}) corresponds to the Langevin sources that are $\delta$-correlated not only in space but also in time \cite{et2015,belyanin2013}.

\subsection{Spontaneous emission from an ensemble of nonequilibrium fermions in a single-mode cavity}

If we assume the populations to be given, the Heisenberg equations for the off-diagonal elements of the density operator can be averaged over the original state of quasiparticles. After averaging, the off-diagonal elements will depend on the field operators, noise operators, and populations $n_{m\mb{k}} $. The operators of populations $\hat{\rho}_{mm\mb{kk}}$ in Eq.~(\ref{Eq:38}) will be replaced by c-numbers: $\hat{\rho}_{mm\mb{kk}} \implies n_{m\mb{k}}$ ; see \cite{vdovin,belyanin2013}.

The structure of Eqs.~(\ref{Eq:38}) and (\ref{Eq:40}) suggests the substitution $\hat{c}_\nu=\hat{c}_{0\nu}(t) e^{-i\omega_\nu t} , \hat{c}_\nu^\dagger=\hat{c}_{0\nu}^\dagger(t) e^{+i\omega_\nu t}$. Here $\hat{c}_{0\nu}(t) $ and $\hat{c}_{0\nu}^\dagger(t)$ are ``slow'' amplitudes in the following sense: $\langle \dot{\hat{c}}_{0\nu}\rangle \ll \omega_\nu\langle \hat{c}_{0\nu}\rangle$; see \cite{vdovin}. Neglecting any inhomogeneous broadening of the resonance line, a steady-state solution of Eq.~(\ref{Eq:38}) for a single-mode cavity is
\begin{equation}
\label{Eq:42}
\hat{\rho}_{21\mb{kk}} \approx \frac{i\tilde{d}_{21}\tilde{D}_\nu}{\hbar}\frac{\hat{c}_{0\nu}e^{-i\omega_\nu t}(n_{1\mb{k}}-n_{2\mb{k}})}{i(\omega_{21}-\omega_\nu)+\gamma_{21\mb{kk}}} + \underaccent{\infty}{\int} \frac{\hat{F}_{\omega;21\mb{kk}}e^{-i\omega t}d\omega }{i(\omega_{21}-\omega)+\gamma_{21\mb{kk}}}.
\end{equation}
Substituting Eq.~(\ref{Eq:42}) into Eq.~(\ref{Eq:40}) we obtain
\begin{align}
\label{Eq:43}
\dot{\hat{c}}_{0\nu} + ( \Gamma_r+\Gamma_\sigma+i\delta\omega+\gamma)\hat{c}_{0\nu} &=  \frac{i\tilde{d}_{21}\tilde{D}^*_\nu}{\hbar} \sum_\mb{k} \underaccent{\infty}{\int} \frac{\hat{F}_{\omega;21\mb{kk}}e^{-i(\omega-\omega_\nu) t}d\omega }{i(\omega_{21}-\omega)+\gamma_{21\mb{kk}}} \nonumber \\ 
&+\underaccent{\infty}{\int} \hat{L}^{(\nu)}_{r\omega'}e^{-i(\omega'-\omega_\nu) t}d\omega' +\underaccent{\infty}{\int} \hat{L}^{(\nu)}_{\sigma\omega''}e^{-i(\omega''-\omega_\nu) t}d\omega'' 
\end{align}
where
\begin{gather}
\label{Eq:44}
\delta\omega=\Omega^2\text{Re}\sum_k\frac{n_{1\mb{k}}-n_{2\mb{k}}}
{(\omega_{21}-\omega_\nu)-i\gamma_{21\mb{kk}}},\hspace{5 mm}
\gamma=\Omega^2\text{Im}\sum_k\frac{n_{1\mb{k}}-n_{2\mb{k}}}
{(\omega_{21}-\omega_\nu)-i\gamma_{21\mb{kk}}},
\\
\label{Eq:45}
\Omega^2=\frac{|\tilde{d_{21}}|^2|\tilde{D_\nu}|^2}{\hbar^2}=
\frac{|\tilde{d_{21}}|^22\pi\omega_\nu}{\hbar L_xL_yG(L_z,\omega_\nu)}.
\end{gather}

The frequency shift $\delta\omega$ of the ``cold'' cavity mode is due to the optical transitions between electron states in a QW. We can redefine the cavity mode frequency assuming that the effect of electrons has been included in $\omega_\nu$ from the very beginning (a ``hot'' cavity mode). The decay rate $\gamma$ describes absorption by electrons; the population inversion corresponds to $\gamma<0$. If $(\gamma+\Gamma_r+\Gamma_\sigma)<0$ the instability develops and the field grows with time; we don't consider this case here. 

The steady-state solution of Eq.~(\ref{Eq:43}) has the form
\begin{multline}
\label{Eq:46}
	\hat{c}_{0\nu} = \frac{i \tilde{d}_{12} \tilde{D}_{\nu}^*}{\hbar}
\sum_{\mb{k}}\int\limits_{\infty}\frac{\hat{F}_{\omega;21\mb{kk}}e^{-i(\omega-\omega_\nu)t}d\omega}
{[i(\omega_\nu-\omega)+\Gamma_r+\Gamma_\sigma+\gamma]\times[i(\omega_{21}-\omega)+\gamma_{21\mb{kk}}]}\\
+\int\limits_{\infty}\frac{\hat{L}^{(\nu)}_{r\omega'}e^{-i(\omega'-\omega_\nu)t}d\omega'}
{[i(\omega_\nu-\omega')+\Gamma+\Gamma_\sigma+\gamma]}
+\int\limits_{\infty}\frac{\hat{L}^{(\nu)}_{\sigma\omega''}e^{-i(\omega''-\omega_\nu)t}d\omega''}
{[i(\omega_\nu-\omega'')+\Gamma+\Gamma_\sigma+\gamma]}.
\end{multline}
Next, we use the Hermitian conjugate of Eq.~(\ref{Eq:46}) to find the value of $\langle\hat{c}_{0\nu}^\dagger\hat{c}_{0\nu}\rangle$, assuming that the statistics of noise operators $\hat{F}_{21\mb{kk}}(t)$, $\hat{L}_r^{(\nu)}(t)$ and $\hat{L}_\sigma^{(\nu)}(t)$ are independent from each other. Using Eqs.~(\ref{Eq:39}) and (\ref{Eq:41}) we obtain
\begin{multline}
\label{Eq:47}
	\langle\hat{c}_{0\nu}^\dagger\hat{c}_{0\nu}\rangle
=\Omega^2\sum_{\mb{k}}\int\limits_{\infty}\frac{d\omega}{\pi}
\frac{\gamma_{21\mb{kk}}n_{2\mb{k}}}{[(\omega_\nu-\omega)^2+(\Gamma_r+\Gamma_\sigma+\gamma)^2]\times[(\omega_{21}-\omega)^2+\gamma^2_{21\mb{kk}}]}\\
+\frac{\Gamma_r}{\Gamma_r+\Gamma_\sigma+\gamma}	n_{T_r}(\omega_\nu)
+\frac{\Gamma_\sigma}{\Gamma_r+\Gamma_\sigma+\gamma}	n_{T_\sigma}(\omega_\nu).
\end{multline}

For simplicity, we neglect the last two terms in Eq. (\ref{Eq:47}) which describe the contribution of the EM background of a surrounding medium and thermal radiation of the material inside a cavity. The power emitted by electrons into the outside space is $P=2\Gamma_r\times\hbar\omega_\nu\times\langle\hat{c}_{0\nu}^\dagger\hat{c}_{0\nu}\rangle$:
\begin{equation}
\label{Eq:48}
P=\hbar\omega_\nu\Omega^2\sum_{\mb{k}}\int\limits_{\infty}\frac{d\omega}{\pi}
\frac{2\Gamma_r\gamma_{21\mb{kk}}n_{2\mb{k}}}{[(\omega_\nu-\omega)^2+(\Gamma_r+\Gamma_\sigma+\gamma)^2]\times[(\omega_{21}-\omega)^2+\gamma^2_{21\mb{kk}}]}.
\end{equation}

Equation (\ref{Eq:48}) for the spontaneous emission power is the main result of this section. It has two obvious limiting cases: 
	
(i) The transition line is much narrower than the cavity resonance: $\Gamma_r+\Gamma_\sigma+\gamma\gg\gamma_{21\mb{kk}}$. In this case we can get from Eq.~(\ref{Eq:48}) 
\begin{equation}
\label{Eq:49}
	P=\hbar\omega_\nu\lbrack A^{(N)}_{2\rightarrow 1}(\Delta \omega)_{\Delta\omega=\Delta\omega^{(1)}_{eff}}\rbrack\cdot
\frac{\Gamma_r}{\Gamma_r+\Gamma_\sigma+\gamma}\cdot
\frac{\Gamma_r+\Gamma_\sigma+\gamma}
{(\omega_\nu-\omega_{21})^2+(\Gamma_r+\Gamma_\sigma+\gamma)^2}\cdot\sum_{\mb{k}}n_{2\mb{k}},
\end{equation}
where $A^{(N)}_{2\rightarrow1}(\Delta\omega)$ is the probability of the spontaneous emission in a cavity given by Eq.~(\ref{Eq:34}) and $\Delta\omega_{eff}^{(1)}=2(\Gamma_r+\Gamma_\sigma+\gamma)$. The second factor in Eq.~(\ref{Eq:49}) determines the fraction of the radiation which escaped outside. The third factor is due to a position of the narrow transition line within a broader cavity mode line. The last factor is a number of radiating particles: $\sum_kn_{2\mb{k}}\Rightarrow \displaystyle \frac{S}{(2\pi)^2}\int n_{2\mb{k}}d^2k$.

(ii) The transition line is much wider than the cavity resonance: $\Gamma_r+\Gamma_\sigma+\gamma\ll\gamma_{21\mb{kk}}$. In this case
\begin{equation}
\label{Eq:50}
		P=\hbar\omega_\nu\lbrack A^{(N)}_{2\rightarrow 1}(\Delta \omega)_{\Delta\omega=\Delta\omega^{(2)}_{eff}}\rbrack\cdot
\frac{\Gamma_r}{\Gamma_r+\Gamma_\sigma+\gamma}\cdot
\sum_{\mb{k}}\frac{\left<\gamma_{21}\right>\gamma_{21\mb{kk}}n_{2\mb{k}}}
{(\omega_\nu-\omega_{21})^2+\gamma_{21\mb{kk}}^2}.
\end{equation}
Instead of the cavity linewidth $2(\Gamma_r+\Gamma_\sigma+\gamma)$ Eq.~(\ref{Eq:50}) contains the homogeneous linewidth $\Delta\omega_{eff}^{(2)}=\left<\gamma_{21}\right>$ where the right-hand side is an average value of $\gamma_{21\mb{kk}}$. Now the third factor is due to a position of the narrow cavity mode line within a broader transition line. Therefore, the effective quality factor is determined by greater of the two values, $\Gamma_r+\Gamma_\sigma+\gamma$ or $\left<\gamma_{21}\right>$. The spontaneous emission efficiency is proportional to the factor $\displaystyle\frac{\Gamma_r}{\Gamma_r+\Gamma_\sigma+\gamma}$, where $\gamma$ is the decay rate of the field due to absorption by electrons. Since $\gamma$ depends on the electron density, the spontaneous emission efficiency per particle also depends on their density. 

One can further simplify Eq.~(\ref{Eq:48}) if relaxation constants $\gamma_{21\mb{kk}}$ do not depend on $\mb{k}$, i.e. $\gamma_{21\mb{kk}} \equiv \gamma_{21}$: 
\begin{equation}
\label{Eq:48a}
P=\hbar\omega_\nu\Omega^2\int\limits_{\infty}\frac{d\omega}{\pi}
\frac{2\Gamma_r\gamma_{21}}{[(\omega_\nu-\omega)^2+(\Gamma_r+\Gamma_\sigma+\gamma)^2]\times[(\omega_{21}-\omega)^2+\gamma^2_{21}]} \sum_{\mb{k}} n_{2\mb{k}}. 
\end{equation}
Here $\gamma$ is defined by Eq.~(\ref{Eq:44}); for $\gamma_{21\mb{kk}} \equiv \gamma_{21}$ it becomes
\begin{equation} 
\label{Eq:44a}
\gamma=\Omega^2 \frac{\gamma_{21}}
{(\omega_{21}-\omega_\nu)^2 + \gamma_{21}^2} \sum_k \left(n_{1\mb{k}}-n_{2\mb{k}}\right),
\end{equation}
where $\Omega^2$ is given by Eq.~(\ref{Eq:45}). Using Eq.~(\ref{Eq:34}), one can rewrite Eq.~(\ref{Eq:48a}) as
\begin{equation}
\label{Eq:48b}
P = \hbar \omega_{\nu} A_{2\to 1}^{(N)} \sum_{\mb{k}} n_{2\mb{k}}, 
\end{equation}
where
\begin{equation}
\label{Eq:34a}
A_{2\to 1}^{(N)}= \frac{2\pi |\tilde{d}_{21}|^2 \left(\displaystyle \frac{4\omega_{21}}{\Delta \omega_{eff}}\right)} {\hbar L_x L_yG(L_z,\omega_\nu)}
\end{equation}
and
\begin{equation}
\label{eff}
\frac{1}{\Delta \omega_{eff}} = \int\limits_{\infty}\frac{d\omega}{4\pi}
\frac{2\Gamma_r\gamma_{21}}{[(\omega_\nu-\omega)^2+(\Gamma_r+\Gamma_\sigma+\gamma)^2]\times[(\omega_{21}-\omega)^2+\gamma^2_{21}]}. 
\end{equation}

For a cavity filled with a uniform and dispersionless medium with dielectric constant $\varepsilon$  one can further simplify Eq.~(\ref{Eq:48b}) as
\begin{equation}
\label{Eq:48c}
P = \left[ \hbar \omega_{\nu} A^{(0)} \sum_{\mb{k}} n_{2\mb{k}} \right] \left[ \frac{6}{\pi^2} \frac{\left( \lambda/2\sqrt{\varepsilon}\right)^3}{L_x L_y L_z} \right] Q_{eff}, 
\end{equation}
Where $\displaystyle A^{(0)} = \frac{4\omega^3|d_{21}|^2 \sqrt{\varepsilon} }{3\hbar c^3}$ is the spontaneous emission rate into free space filled with dielectric medium $\varepsilon$ and $Q_{eff} = \displaystyle \frac{\omega_{21}}{\Delta \omega_{eff}}$ is the effective quality factor. The term in the first brackets on the rhs of Eq.~(\ref{Eq:48c}) is the power of spontaneous emission into free space; the term in the second brackets is the geometric enhancement due to a subwavelength cavity. 

The integral in Eq.~(\ref{eff}) is a product of two Lorentzians which can be easily evaluated analytically but is a bit cumbersome. Assuming for simplicity exact resonance between the transition frequency and the cavity resonance,  $\omega_{\nu} = \omega_{21}$, we obtain 
\begin{equation}
\label{eff2} 
Q_{eff} = \frac{\omega_{21} \Gamma_r}{2(\Gamma_r+\Gamma_\sigma+\gamma)(\gamma_{21} + \Gamma_r+\Gamma_\sigma+\gamma)} \rightarrow \frac{\omega_{21}}{2(\gamma_{21} + \Gamma_r)},
\end{equation}
where the last expression is in the limit $\Gamma_r \gg \Gamma_{\sigma} + \gamma$.  

For a fixed transition linewidth $\gamma_{21}$ we normalize $Q_{eff}$ by the Q-factor of the radiative transition $\displaystyle \frac{\omega_{21}}{2\gamma_{21}}$ and plot the normalized Q-factor  $Q_{norm} = \displaystyle \frac{2\gamma_{21}}{\Delta \omega_{eff}}$ as a function of the cavity linewidth $\Gamma_r$; see Fig.~2. As shown in Fig.~2, it makes no sense to increase the Q-factor of the cavity mode $\displaystyle \frac{\omega_{\nu}}{2\Gamma_r}$ beyond the value determined by $\Gamma_r \sim \gamma_{21}$ when the effective Q-factor $Q_{eff} $ reaches its maximum value $\displaystyle \sim 5 \frac{\omega_{\nu}}{2\gamma_{21}}$. For smaller values of $\Gamma_r$  the intracavity quantum efficiency will stay roughly the same, limited by the dissipation rate $\gamma_{21}$ of the optical polarization, whereas the radiation power outcoupled from the cavity reduces $\propto \Gamma_r$. 

\begin{figure}[h]
    \centering
    \includegraphics[width=0.8\columnwidth]{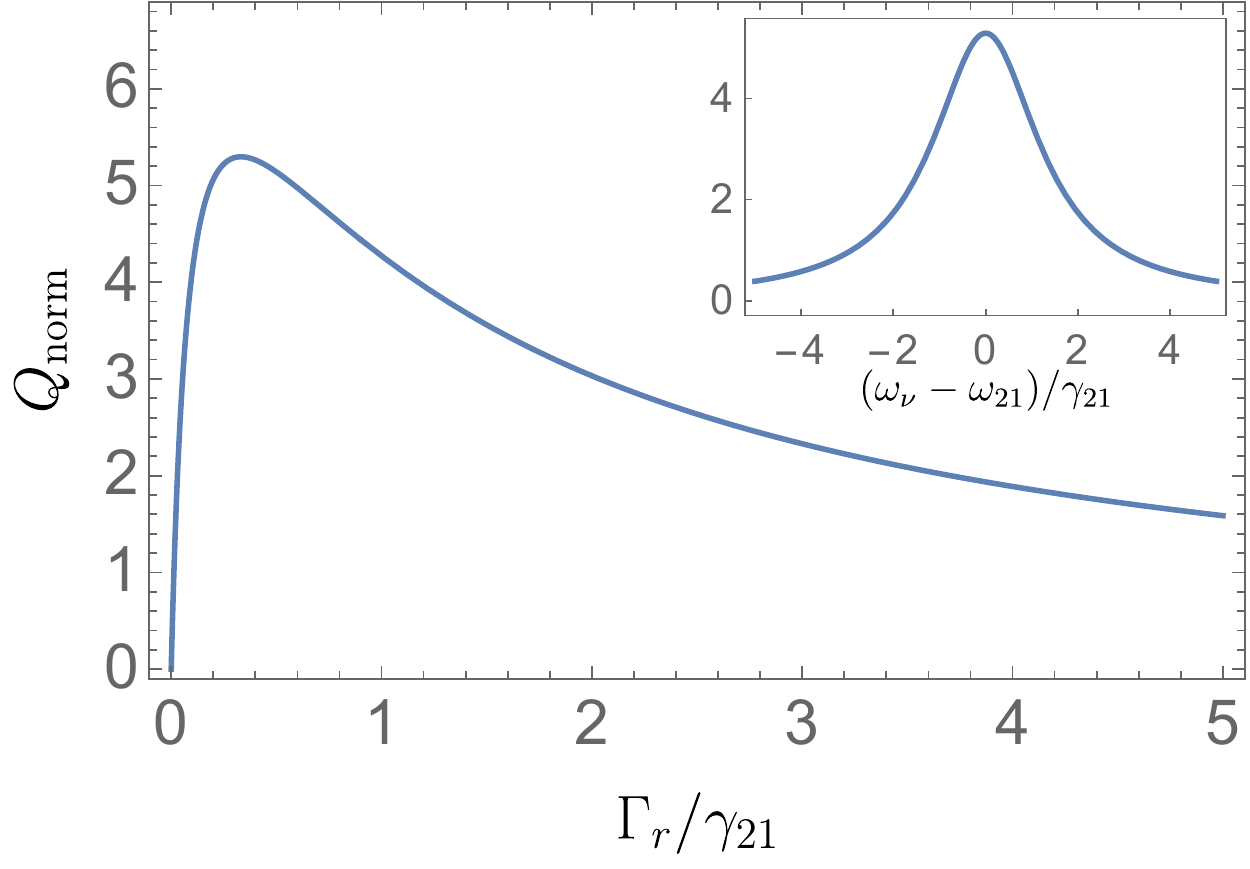}
    \caption{The normalized effective Q-factor as a function of the normailzed cavity linewidth $\Gamma_r / \gamma_{21}$ at exact resonance $\omega_{21} = \omega_{\nu}$. Inset: the normalized effective Q-factor as a function of frequency detuning at $\Gamma_r = \gamma_{21}$.  }
	\label{fig2}
\end{figure}

For mid-infrared intersubband transitions in multiple QW nanocavities at $\hbar \omega_{21} \sim 100-200$ meV and full linewidth $2 \gamma_{21} = 10$ meV \cite{lee2016}  the maximum $Q_{eff} \sim 50-100$ and the geometric enhancement in Eq.~(\ref{Eq:48c}) can add another factor of $10-100$. For THz intersubband transitions $Q_{eff} $ is similar whereas the geometric enhancement is a factor of 10 higher. For a near-infrared transition in semiconductor QWs or 2D semiconductors the frequency is $ \sim 5-10$ times higher, but the linewidth is $2-3$ times higher as well, so $Q_{eff}$ can be about 100-300. This example also suggests that an optimal radiative loss from a cavity (or a cavity mode linewidth) for semiconductor 2D emitters should be of the order of 5-10 meV. 

All results in this section are applicable to a waveguide at the cutoff frequency.

In conclusion, using consistent Heisenberg-Langevin approach we derived general analytic formulas describing the spontaneous emission of 2D emitters placed in plane-parallel subwavelength cavities or waveguides. We found that a significant enhancement of the outcoupled spontaneous emission and quantum efficiency of semiconductor quantum devices can be achieved for realistic device parameters. 

This material is based upon work supported by the Air Force Office of Scientific Research under award numbers FA9550-15-1-0153, FA9550-17-1-0341, and FA9550-14-1-0376.  M.T. acknowledges the support from RFBR grant No.~17-02-00387 and Ministry of Education Science of the Russian Federation contract No.~14.W03.31.0032.

\appendix

\section{EM field quantization in a subwavelength cavity filled with a layered dispersive medium \label{appendix:EM quantization}}

We start from the expression for the energy of a classical EM field in a nonmagnetic medium \cite{LL,fain}: 
\begin{equation}
\label{Eq:A1}
W=\frac{B^2}{8\pi}+ \frac{1}{4\pi}  \int\limits_C^t \mb{E}\dot{\mb{D}}dt.
\end{equation}
According to  Eq.~(\ref{Eq:12}) in our case the electric field and electric induction vectors are equal to 
\begin{equation}
\label{Eq:A2}
\mb{E} =\mb{z_0}D_\nu \frac{\zeta_\nu(x,y)}{\varepsilon(\omega_\nu,z)}e^{-i\omega_\nu t} +\mathrm{C.C.} ,
\hspace{.4 cm} 
\mb{D}=\mb{z_0}D_\nu \zeta_\nu(x,y)e^{-i\omega_\nu t} +\mathrm{C.C.}
\end{equation}
For a non-uniform medium with frequency dispersion the spatial distribution of the field depends explicitly on the frequency $\omega_\nu$; this fact requires certain modification of the approach used in \cite{fain,tokman2015} to calculate the field energy $W$. Assume an adiabatically slow ``turning on'' of the electric induction at the moment of time $t = C$, i.e. $D_\nu \implies D_\nu(t), D_\nu(C)=0, \dot{D_\nu}\ll \omega_\nu D_\nu $. In this case one can write 

\begin{equation}
\label{Eq:A3}
\left.
\begin{matrix}
\dot{\mb{D}}= \mb{z_0}\zeta_\nu(x,y)e^{-i\omega_\nu t} (-i\omega_\nu D_\nu +\dot{D_\nu}) +\mathrm{C.C.}\\\
\\
\mb{E} \approx  \mb{z_0} \zeta_\nu(x,y)e^{-i\omega_\nu t} \left( \displaystyle \frac{D_\nu}{\varepsilon(z,\omega_\nu)} +i\dot{D}_\nu  \frac{\partial}{\partial\omega}\left(\frac{1}{\varepsilon(z,\omega)}\right)_{\omega=\omega_\nu} \right)  +\mathrm{C.C.}
\end{matrix}
\right\}.
\end{equation}
In addition, we take into account that for monochromatic fields $\mb{E}=\mb{E}_\nu(\mb{r})e^{-i\omega_\nu t} +\mathrm{C.C.}$ ,  $\mb{B}=\mb{B}_\nu(\mb{r})e^{-i\omega_\nu t} + \mathrm{C.C.}$  and $\mb{D}=\mb{D}_\nu(\mb{r})e^{-i\omega_\nu t} + \mathrm{C.C.}$  in a cavity or under periodic boundary conditions the flux of the complex vector $\mb{E}_\nu \times \mb{B}^*_\nu$  through a surface enclosing volume   is equal to zero. This allows one to prove that (see also \cite{vdovin,tokman2016}) 

\begin{equation}
\label{Eq:A4}
\int\limits_V \mb{B}_\nu \mb{B}^*_\nu d^3r=\int\limits_V \mb{D}_\nu \mb{E}^*_\nu d^3r,
\end{equation}
Using  Eqs.~(\ref{Eq:A1}) -~(\ref{Eq:A4}) one can get

\begin{equation} 
\int\limits_V W d^3r = \frac{|D_\nu|^2}{4\pi} \int\limits_S \zeta_\nu \zeta^*_\nu d^2r \times \int\limits_{-\frac{L_z}{2}}^{+\frac{L_z}{2}} \left[\frac{2}{\varepsilon(z,\omega_\nu)}-\omega_\nu \frac{\partial}{\partial\omega}\left(\frac{1}{\varepsilon(z,\omega)}\right)_{\omega=\omega_\nu} \right]dz . \nonumber
\end{equation}
After we impose the requirement $\int_V Wd^3r = \hbar \omega_\nu$ and take into account the relation

\begin{equation}
\frac{2}{\varepsilon}- \omega \frac{\partial}{\partial \omega}\left(\frac{1}{\varepsilon}\right)= \frac{1}{\varepsilon^2\omega}\frac{\partial(\omega^2 \varepsilon)}{\partial\omega} \nonumber
\end{equation}
we arrive at the normalization condition  Eq.~(\ref{Eq:14}).

\section{Matrix elements of the interaction Hamiltonian for fermions coupled to an EM field in a cavity or a waveguide  \label{appendix:Matrix elements}}

The explicit form of the matrix elements in  Eq.~(\ref{Eq:22}) is \\
 (i) in the waveguide:
\begin{equation}
\label{Eq:B1}
\zeta^{(q_x)}_{\mb{k'}\mb{k}} = \delta_{k'_x , k_x+q_x} Y_{k'_y,k_y} ,
\end{equation}
where
\begin{equation}
Y_{k'_y,k_y} =  \displaystyle \frac{\sin\left[\left(k_y + \frac{\pi}{L_y}-k'_y\right)\frac{L_y}{2}\right]}{\left(k_y + \frac{ \pi}{L_y}-k'_y\right)L_y} +\frac{\sin\left[\left(k'_y + \frac{\pi}{L_y}-k_y\right)\frac{L_y}{2}\right]}{\left(k'_y + \frac{\pi}{L_y}-k_y\right)L_y} ; \nonumber
\end{equation}
(ii) in the cavity:
\begin{equation}
\label{Eq:B2}
\zeta^{(N)}_{\mb{k'}\mb{k}} = Y_{k'_y,k_y} X_{k'_x,k_x},
\end{equation}
where
\begin{equation}
\begin{matrix}
 X^{(odd)}_{k'_x,k_x} = \displaystyle \frac{\sin\left[\left(k_x + \frac{N_{odd} \pi}{L_x}-k'_x\right)\frac{L_x}{2}\right]}{\left(k_x + \frac{N_{odd} \pi}{L_x}-k'_x\right)L_x} +\frac{\sin\left[\left(k'_x + \frac{N_{odd} \pi}{L_x}-k_x\right)\frac{L_x}{2}\right]}{\left(k'_x + \frac{N_{odd} \pi}{L_x}-k_x\right)L_x} ,\\\
 
 \\\
  X^{(even)}_{k'_x,k_x} = i \displaystyle \frac{\sin\left[\left(k'_x + \frac{N_{even} \pi}{L_x}-k_x\right)\frac{L_x}{2}\right]}{\left(k'_x + \frac{N_{even} \pi}{L_x}-k_x\right)L_x} - i\frac{\sin\left[\left(k_x + \frac{N_{even} \pi}{L_x}-k'_x\right)\frac{L_x}{2}\right]}{\left(k_x + \frac{N_{even} \pi}{L_x}-k'_x\right)L_x}.
\end{matrix} \nonumber 
\end{equation}
These expressions are presented in the form which shows explicitly the factors of the type $ \displaystyle  \frac{\sin(Ax)}{x}$.

When calculating the radiated power by an ensemble of fermions we need to know the squares of matrix elements summed over electron $\mb{k}$-states, in particular $\sum_{k'_y}Y_{k'_y,k_y} Y_{k_y,k'_y}$ and $
\sum_{k'_x}  X_{k'_x,k_x} X_{k_x,k'_x}$. Taking into account that 

\begin{equation}
\int\limits_{-\infty}^{+\infty} \frac{\sin^2x}{x^2} dx = \pi ,
\hspace{4mm}
\int\limits_{-\infty}^{+\infty} \frac{\cos^2x}{\left(\frac{\pi}{2}\right)^2 - x^2}dx = 0 , \nonumber
\end{equation}
we obtain
\begin{equation}
\label{Eq:B3}
  \sum_{k'_y}Y_{k'_y,k_y} Y_{k_y,k'_y} \implies \frac{L_y}{2\pi} \int\limits_\infty Y_{k'_y,k_y} Y_{k_y,k'_y} dk'_y =\frac{1}{2}, 
  \hspace{0.2cm}
    \sum_{k'_x}X_{k'_x,k_x} X_{k_x,k'_x} \implies \frac{L_x}{2\pi} \int\limits_\infty X_{k'_x,k_x} X_{k_x,k'_x} dk'_x = \frac{1}{2} .
\end{equation}
Since  $\int_S \zeta_{q_x} \zeta^*_{q_x} d^2r =S/2 $ and $\int_S \zeta_N \zeta^*_N d^2r =S/4$ ,  Eq.~(\ref{Eq:B3}) give the equation $\sum_{\mb{k}'}  \zeta_{\mb{k}'\mb{k}}^{(\nu)} \zeta_{\mb{k}\mb{k}'}^{(\nu)  \dagger} = S^{-1} \int_S \zeta_\nu \zeta_\nu^* d^2r$, which is used in Sec.~IIb.

\section{Commutation relations for Langevin sources  \label{appendix:Commutation relations}}

Consider a quantum oscillator described by the Hamiltonian $\hat{H}= \hbar \omega(\hat{c}^\dagger \hat{c} + 1/2) $. After substituting $\hat{c}= \hat{c}_0 e^{-i\omega t}$ and $\hat{c}^\dagger= \hat{c}^\dagger_0 e^{-i\omega t}$  the Heisenberg equations of motion take the form $\dot{\hat{c}}_0 = 0$, $\dot{\hat{c}}^\dagger_0 = 0$. The simplest model of interaction with a dissipative reservoir modifies these equations as follows: $\dot{\hat{c}}_0  + \Gamma \hat{c}_0 =0$, $\dot{\hat{c}}^\dagger_0  + \Gamma \hat{c}^\dagger_0 =0$. However, this modification leads to violation of boson commutation relation $[\hat{c}_0 ,\hat{c}^\dagger_0]=1$. To resolve this issue and preserve the commutator one has to add the Langevin sources to the right-hand side of Heisenberg equations \cite{SZ}:

\begin{equation}
\label{Eq:C1}
\dot{\hat{c}}_0  + \Gamma \hat{c}_0 = \hat{L},
\hspace{.4cm}
\dot{\hat{c}}^\dagger_0  + \Gamma \hat{c}^\dagger_0 = \hat{L}^\dagger.
\end{equation}
Langevin noise operators in  Eq.~(\ref{Eq:C1}) describe fluctuations in a dissipative system. Note that $\langle\hat{L}\rangle=0$ ; the notation $\langle\cdots\rangle$ means averaging over the statistics of the dissipative reservoir and over the initial quantum state $|\Psi \rangle$ within the Heisenberg picture.

The operator $\hat{L}$ is usually defined together with the relaxation constant $\Gamma$ within a given model of the reservoir \cite{SZ}. However, the commutation relations for a noise operator can be obtained directly from the given form of the relaxation operator if we require that standard commutation relations $[\hat{c}_0 ,\hat{c}^\dagger_0]=1, [\hat{c}_0 ,\hat{c}_0]=1$,   be satisfied at any moment of time. Indeed, let’s substitute the solution of the operator-valued equations~(\ref{Eq:C1})

\begin{equation}
\label{Eq:C2}
\hat{c}_0 = \hat{c}_0(0) e^{-\Gamma t} + \int\limits_0^t e^{ \Gamma (t'-t) } \hat{L}(t') dt',
\hspace{.4cm}
\hat{c}^\dagger_0 = \hat{c}^\dagger_0(0) e^{-\Gamma t} + \int\limits_0^t  e^{ \Gamma (t'-t) } \hat{L}^\dagger(t') dt'
\end{equation}
into the commutators. It is easy to see that the standard commutation relations will be satisfied if, first of all, the field operators at an initial moment of time, $\hat{c}_0 (0)$ and $\hat{c}^\dagger_0(0)$, commute with Langevin operators $\hat{L}(t)$ and $\hat{L}^\dagger(t)$ in any combination. Second, the following condition has to be satisfied:  

\begin{equation}
\label{Eq:C3}
[\hat{L} ,\hat{c}^\dagger_0]=[\hat{c}_0,\hat{L}^\dagger] = \Gamma.
\end{equation}
Substituting  Eq.~(\ref{Eq:C2}) into  Eq.~(\ref{Eq:C3}) and using the identity $\int\limits_0^t  X(t') \delta(t-t') dt' = X(t)/2$  we arrive at
\begin{equation}
\label{Eq:C4}
[\hat{L}(t') , \hat{L}^\dagger(t)] = 2\Gamma \delta(t-t') .
\end{equation}

\end{document}